# Memories of Julian Schwinger

Edward Gerjuoy*

## ABSTRACT

The career and accomplishments of Julian Schwinger, who shared the Nobel Prize for physics in 1965, have been reviewed in numerous books and articles.  For this reason these Memories, which seek to convey a sense of Schwinger's remarkable talents as a physicist, concentrate primarily (though not entirely) on heretofore unpublished pertinent recollections of the youthful Schwinger by this writer, who first encountered Schwinger in 1934 when they both were undergraduates at the City College of New York.

*University of Pittsburgh Dept. of Physics and Astronomy, Pittsburgh, PA 15260.

# Memories of Julian Schwinger

Julian Seymour Schwinger, who shared the 1965 Nobel prize in physics with Richard Feynman and Sin-itiro Tomonaga, died in Los Angeles on July 16, 1994 at the age of 76. He was a remarkably talented theoretical physicist, who—through his own publications and through the students he trained—had an enormous influence on the evolution of physics after World War II. Accordingly, his life and work have been reviewed in a number of publications [1-3, inter alia]. There also have been numerous published obituaries, of course [4-5, e.g.]

These Memories largely are the text of the Schwinger portion of a colloquium talk, "Recollections of Oppenheimer and Schwinger", which the writer has given at a number of universities, and which can be found on the web [6]. The portion of the talk devoted to J. Robert Oppenheimer—who is enduringly famous as director of the Los Alamos laboratory during World War II—has been published [7], but the portion devoted to Julian is essentially unpublished; references 7-9 describe the basically trivial exceptions to this last assertion. The aforesaid colloquium talk linked Oppenheimer and Schwinger because for a time during the years immediately preceding Pearl Harbor Schwinger served as Oppenheimer's post doc.

However my special motivation for devoting much of my talk to my recollections of Schwinger, and for writing these Memories therefore, has been the impression—gained from many conversations with physics graduate students during the years following his death in 1994—that Schwinger's fame and reputation slowly were being forgotten although the fame and reputation of his fellow Nobel Laureate Richard Feynman have continued unabated, perhaps even increasing, since Feynman's death in 1988.

A Panel distributed for posting by the American Physical Society illustrates what I have just asserted [10]. The right side of the Panel is a portrait of Richard Feynman. A portion of the text to the left of Feynman's photo says: "Together with American colleagues and Japanese physicists who had worked along similar lines while they were out of touch with the West during the war, FEYNMAN solved the problem by creating Quantum Electrodynamics (QED)."

I ask (I'm sure many of my readers will ask): Where is Schwinger's portrait? Indeed where even is Schwinger's name? After all this is a Panel officially created by the American Physical Society (APS), who certainly knew that Feynman shared with Schwinger the 1965 physics Nobel Prize for the development of QED. In fact, Schwinger performed his calculations explaining the two leading QED experiments, namely the Lamb Shift and the value of the electron anomalous magnetic moment, before Feynman did. It's true that if you look carefully at the entire Panel there is a small box in the middle of the Panel which does mention Schwinger, but there's no photograph and believe me this mention of Schwinger is not easy to find. This downgrading of Schwinger by the APS is puzzling not only because Schwinger was the first formulator of modern QED, but also because although they were both born in 1918 only three months apart, Schwinger began to publish far ahead of Feynman, and indeed was much more famous than Feynman until about 1955.

In fact here's what Feynman himself had to say about his first meeting with Julian, when they both were 25 years old [11]:

"It was not until I went to Los Alamos that I got a chance to meet Schwinger. He had already a great reputation because he had done so much work...and I was very anxious to see what this man was like. I'd always thought he was older than I was because he had done so much more. At the time I hadn't done anything. And he came and gave us lectures. I believe they were on nuclear physics. I'm not sure exactly the subject, but it was a scene you probably all have seen once. The beauty of one of his lectures. He comes in, with his head a little bit to one side. He comes in like a bull into a ring and puts his notebook down and then begins. And the beautiful, organized way of putting one idea after the other. Everything very clear from the beginning to the end…I was supposed to be a good lecturer according to some people, but this was really a masterpiece… So I was very impressed, and the times I got then to talk to him, I learned more."

## SCHWINGER'S EARLY LIFE AND CAREER

So let me now give you a brief summary of Schwinger's early life and career. Like Feynman, Schwinger was born and brought up in New York City. Unlike most other physicists of his generation, or generations thereafter for that matter, he very early became focused on physics [12]. Schwinger himself, in autobiographical material submitted to the Nobel Foundation, indicated that "the principal direction of his life was fixed at an early age by an intense awareness of physics, and its study became an all-engrossing activity" [13].

Moreover Schwinger learned physics essentially completely on his own, with little or no help from the formal educational system. He began this self-education at a very young age, primarily by reading books in the public library, and carried it on so effectively that by age 17 he was doing calculations with more established physicists which merited publication as Letters to the Editor of the Physical Review [14]; by this time he already had established his habit of working through the night, about which more later. In 1937, at age 19, he published, on his own, two full scale papers in the Physical Review, as well as a Letter to the Editor [15]; that same year he also published, with Edward Teller, a Letter to the Editor and a full scale paper, both on the scattering of neutrons by ortho- and para-hydrogen [16].

This writer first met Julian in 1934, shortly after Julian entered the City College of New York (CCNY). We met when we both took "Intermediate mechanics," the first course in mechanics after elementary physics. I never knew him or had heard of him before. Everybody but the course instructor, a drudge, immediately recognized Julian as another species. Most of we students had just learned how to take vector products, and here Schwinger was performing manipulations like making integrals vanish using the divergence theorem. Note there is no doubt that at the time CCNY had the most outstanding student body in the U.S. But the faculty, with a very few exceptions, just didn't match up.

So now I can tell you my main claim to fame. At the end of the semester I got an A, but Schwinger only got a B. Julian, because of his habit of working at night, hardly ever came to class, except to take the exams. So the instructor disliked Schwinger, but he couldn't give him a really bad grade because of Julian's exam scores. Hence the B.

Understand that at the time the CCNY curriculum had a huge, and I mean huge, number of required courses outside the major. This writer took: four semesters of public speaking; a semester of ancient history; a semester of modern history; a semester of economics; four semesters of English literature; a year of chemistry; a year of biology; a semester of engineering drafting, etc., etc.

Thus, since Julian hardly ever came to class, and couldn't pass the exams without coming to class in subjects other than math and physics, by the summer of 1935 Schwinger essentially was flunking out of CCNY. Fortunately, this disaster began to loom just about when he was in the process of completing his Letter to the Editor Physical Review written with Lloyd Motz [14], who happened to be both an instructor at CCNY and a Ph.D. student of Isidore Rabi (recipient of the 1944 physics Nobel Prize) at Columbia University. By extremely lucky chance this collaboration with Motz brought Schwinger to the attention of Rabi. Rabi has described this meeting with Schwinger as follows [17]:

"There was this paper of Einstein, Podolsky and Rosen, and I was reading it. And one of my ways of trying to understand something was to call in a student, explain it to him and then argue about it. And [in this case] this was Lloyd Motz who was also at that time, I think an instructor or something at City College.. And then he said there was somebody waiting for him outside, so I said call him in. So he called him in and there was a young boy there. [I] asked him to sit down and we continued. And then at one point there was a bit of an impasse and this kid spoke up and used the completeness theorem...to settle an argument...I was startled. What's this, what's this? So then I wanted to talk to him which I did...It turned out I was told he was having difficulty at City College. I knew the people he was having difficulty with and it's an honorable thing to have difficulty with those people...I suggested transferring to Columbia and then got a transcript from City College, and took it to one of the officials—I forget who it was. Now what about a fellowship, scholarship for this guy? He looked at the transcript and said, on the basis of this we wouldn't admit him. I [then] said something very tactless—I said, suppose he were a football player...Still the problem remained...[but] I just simply overrode them and he was admitted."

It is difficult to even speculate on what would have been Schwinger's fate had he not come to Rabi's attention. At Columbia, Schwinger was permitted to take whatever courses he liked. He was awarded his B.S. in 1936, by which time he already had done enough work to warrant a Ph.D. He continued his research as a student of Rabi's, until actually awarded his Ph.D. in 1939, at age 21.

## SCHWINGER AS J. ROBERT OPPENHEIMER'S POSTDOC

After Julian left CCNY for Columbia. I—who remained at CCNY—largely lost touch with him until the fall of 1939, when (after receiving his Ph.D.) he came to Berkeley as J. Robert Oppenheimer's research associate, a position we now would term "postdoc"; he replaced Leonard Schiff, later the author of a very popular English language quantum mechanics textbook. It was also one of the first [18]. I had come to Berkeley in 1938, as a graduate student hoping to receive my Ph.D. degree from Oppie (as all Oppenheimer's students, and most of his colleagues, referred to Oppenheimer in those pre-WWII days).

Elsewhere [7] I have explained how difficult it was to give a seminar with Oppie in the audience, because of Oppie's ruthless unwillingness to stop putting searching questions to the speaker [19].  Indeed on several occasions he reduced Schiff—who as Oppie's postdoc had to present many many seminars—almost to tears.  Hence once Schwinger arrived in Berkeley, essentially all the students in Oppie's group (who knew nothing about Schwinger) were eagerly anticipating Julian's first seminar, wondering how long it would take Julian to shrivel under Oppie's questioning.  I on the other hand, who not only knew Julian from City College but also had friends at Columbia who informed me how Schwinger had been performing for Rabi, was wondering how Oppie would react to Julian's refusal to shrivel.

And indeed Julian's first seminar in Oppie's group went exactly as I had anticipated.  I no longer recall the subject of the seminar, but I do distinctly remember that Oppie asked Julian question after question, and that Julian answered them all.  At no time did Schwinger show any sign of distress.  Finally Oppie stopped asking questions and allowed Schwinger to proceed uninterrupted.  Moreover, thereafter Oppie never unduly interrupted Julian with questions during any of Julian's lectures.

Rabi relates that Oppie initially was terribly disappointed with Schwinger [20].  Oppie disliked Julian's habit of working at night.  He even thought of asking Julian to go elsewhere.   But he soon learned to respect Julian.  In fact Oppie and Julian co-authored two publications.

## WHAT WORKING WITH JULIAN WAS LIKE

When Julian arrived in Berkeley mine was one of the very few faces he recognized.  Moreover Julian and I, like Julian and Feynman, also were the same age, born just a few months apart.  So it was only natural that Julian and I soon became good friends, much better friends than we had been at City College in fact.  Moreover although Julian did all the teaching of physics between us, I did have a few things to teach Julian.  For instance, I taught Julian to play pool, which he never previously had tried (his talents at theoretical physics did not extend to talents with the cue stick).  And about a year after Julian arrived at Berkeley we started working together on a problem that eventually became one of the publications which constituted my Ph. D. thesis.

In a moment I'll tell you a bit more about what this problem involved and how it originated.  But first let me describe what working with Julian was like.  The most significant factor was the need for me to accommodate to Julian's established habit of working at night.  Of course on some days Julian could not work all night because he had to wake up early enough to go to the physics dept, talk to Oppie, attend or give a seminar, etc. But on by far the most days, here is what my life was like in the year or so I was researching with him:

(i) At 11:45 PM I would meet Julian at the Berkeley campus International House, where he resided during his entire Berkeley period.

(ii) We then would drive to an all-night bistro (Julian had an expensive new car), where Julian would have breakfast.

(iii) After Julian's breakfast (in which I sometimes participated halfheartedly), we drove to the Berkeley physics department's building Leconte Hall, wherein Julian's office was located.

(iv) Once Julian and I reached his office, we then (starting typically about 12:45 AM) put in about three hours of work, after which we would drive to some other bistro, where Julian could have lunch.

(v) After lunch, it was back to Leconte Hall till 8:00 AM, when I had to leave to attend the courses I still was taking (e.g., advanced quantum mechanics as taught by Oppie), as well as to perform my departmental Teaching Assistant duties (a TA on which my personal finances completely depended).

Moreover at some time between 8:00 AM and 11:45 PM I—who had married in 1940, just about when I started working with Julian—had to squeeze in my husbandly duties.   I suppose I must have gotten some sleep during that year working with Julian, but I sure can't remember doing it.

### THE RICHMAN-PETERS RESEARCH PROBLEM

Which brings me to another illustration of Schwinger's awesome talents, all as personally observed by me.  About 5:00 PM one day, shortly after I had begun working with Julian, he and I were in Oppie's office [21], along with Oppie himself, when in walked Chaim Richman and Bernard Peters, two of Oppie's students who were about one year behind me.  They asked Oppie for a publishable research problem, telling him that they (Richman and Peters) had decided to work together.  Oppie thought for a while, and then suggested they work out the cross section for the disintegration of the deuteron by impinging fast electrons; the point was that the fast electrons could exert electromagnetic forces, but no nuclear forces, on the deuteron, so that comparing the cross sections for electron and proton disintegration of the deuteron might be revealing. After a few questions, Richman and Peters departed, assuring Oppie they were going to work on the problem.

By showing up in Oppie's office at 5:00 PM, Schwinger evidenced that he had arisen unusually early that day, but this deviation from his routine did not mean Schwinger was not intending to work as usual that night. So that same night, at the usual 11:45 PM, I met Julian at the International House, all in accordance with the routine described above. I found him at an unusual spot, namely the International House's Western Union desk (ah those happy days of old when people still sent telegrams!) where he was scribbling away on blank Western Union forms. He explained that he had been taken with the problem Oppie had given Richman and Peters, and so—while waiting for me to show up—had decided to work out the cross section for himself. When I arrived, possibly a few minutes later than usual (I no longer remember) Julian told me he was almost finished; indeed—after just a few more minutes at most—he stuffed in his pocket the Western Union form on which he had last scribbled, and off we went to Julian's breakfast.

About six months later both Julian and I happened to again find ourselves in Oppie's office at about 5:00 PM, again along with Oppie, when in trooped Peters and Richman, literally beaming. They told Oppie they had solved the research problem he had given them. Oppie responded, "Let's see what the cross section looks like. Write it on the board." So Peters (or was it Richman?) wrote on the blackboard the cross section, whose expression was far from simple and quite lengthy, taking up a good two boards.  Oppie looked at it, made a few comments, but said the cross section looked correct, when he suddenly turned to Schwinger.  "Julian, didn't you tell me you had worked out the cross section?"

Julian, surprised, started looking through his pockets; astonishingly he turned up that Western Union form he had stuffed away  about six months previously.  It was almost illegible, of course, but Julian was able to decipher it and compare the expression on the board with his result (yes, in the few minutes while waiting to depart for breakfast he had worked out, without any books or notes, just using what was in his head, the cross section for the disintegration of the deuteron by fast electrons!).  Finally Julian said, "Their result looks right, except that I think it's missing a factor of two".  Immediately Oppie, without asking Schwinger for any details of his calculation—therewith showing how much he had learned to respect Julian's talents since those early days when he had complained to Rabi about Julian—commanded Peters and Richman: "Find that factor of two!"

So poor Peters and Richman, who so happily had entered Oppie's office, departed the office with heads bowed, certainly no longer beaming.  In due course they told Oppie they had found the factor of two and published the cross section [22].   But neither of them had a career as a theoretical physicist.  Instead, they each became experimenters (a fact whose implications I will not pursue).

## MY RESEARCH WITH SCHWINGER

I now proceed to describe the problem on which I worked with Schwinger. First let met me explain how the problem originated. When Rabi made his unexpected discovery that the deuteron had a quadrupole moment, Schwinger realized immediately—and I believe was the first to do so—that this fact implied the nuclear two-body interaction had to include non-central so-called "tensor" forces. Since all the earlier two-nucleon system calculations (calculations that had been carried out so tediously and painstakingly by Breit and his co-workers) had ignored tensor forces, it was necessary to completely redo those older calculations.

As it happened, after Schwinger arrived in Berkeley William Rarita came to Oppenheimer's group, on a sabbatical from the Brooklyn College physics department in New York City. Rarita, who to my knowledge never had done any significant research on his own, surely already had heard about Julian through his New York academic connections. So it's not surprising that Julian got Rarita to grind out—under Julian's direction—those needed two-nucleon calculations, i.e., to compute, as a function of the ranges and strengths of the various interaction potentials in the two-nucleon Hamiltonian (interaction potentials still explicitly velocity-independent), quantities such as the binding energy and quadrupole moment of the deuteron, the neutron-proton scattering cross section, the cross section for photo-disintegration of the deuteron, etc. And when I say "grind out" I mean grind out; this still was in the pre-World War II dark ages. So Rarita's assigned task, to which he faithfully dedicated himself, was to pound the best Berkeley physics department Marchant calculator all night long, night after night (since he, like me, was working with Julian) carrying out the calculations Schwinger set before him.

By the time I started working with Julian, it had become pretty clear Rarita's calculations would confirm that the two-nucleon system could be understood in terms of velocity-independent tensor forces. So Julian assigned me the problem of ascertaining whether the next simplest nuclear systems— namely the three-nucleon and four-nucleon systems of hydrogen 3 (the triton), helium 3 and standard helium 4—could also be understood in terms of velocity-independent tensor forces. I don't want to get into the details of those calculations of mine, whose results were published [23], but those calculations did provide me with another demonstration of Schwinger's remarkable computational powers.

The calculations involved evaluating a number of non-trivial matrix elements. Each matrix element involved Pauli spin operations, sometimes quite complicated, followed by angular integrations, in order to reduce the matrix element to an integral over spatial coordinates only. It was these integrals that I had the responsibility of evaluating, either numerically or in closed form, depending on assumptions about the spatial behavior of the nuclear forces. In order to be sure I was evaluating the correct spatial integrals, Julian and I agreed to do the spin and angular operations independently, and then—if our results disagreed—to do the calculations jointly on the blackboard. We disagreed on about 15 out of 100 or so matrix elements. What absolutely infuriated me, infuriates me still in fact, is that every time we disagreed Julian turned out to be right! I really was, perhaps still am, a pretty good theorist, but Julian just was another species. I emphasize that Julian never gloated or put me down. Nor did he do so to anybody else I know of who worked with him.  Julian also knew he was another species.

Before closing this subsection, I should say that once the writing of our paper was finished, I never again worked closely with Julian, but we remained good friends.  For instance I always spent considerable time with Julian whenever I was in his vicinity, e.g., whenever we both were at APS meetings.

## SCHWINGER AT HARVARD

During most of the US WWII years Julian worked at the MIT Radiation Laboratory (Rad Lab), where he was a sensation [24].  He later told me that Oppie wanted him to come to Los Alamos, but that Julian had declined, fearing Oppie would try to dominate him.  In 1945, immediately after the war ended, he accepted a position as Associate Professor at Harvard after being courted by many major universities.  In 1947, at age 29, Harvard promoted him to full professor (a record at the time, surely).

So began the happiest period of Schwinger's life, during which he did the work that earned him the Nobel Prize.   At Harvard he gave his famous series of courses on special topics in theoretical physics, many of which contained all sorts of new results Schwinger never bothered to publish, although some of these course notes were written up and published by Schwinger students.  In particular, concerning Schwinger's nuclear physics lectures, written up by John Blatt, Herman Feshbach has stated [25]:

"[These lecture notes] form an excellent introduction to the applications of quantum mechanics, developing a number of elegant methods of wide applicability. They contain many results specifically important for nuclear physics, many of which were never published or were later rediscovered…It is difficult to exaggerate the impact of these lectures on the generation of physics graduate students in the late forties and fifties by which time a substantial fraction of the notes had been incorporated into the general background material all practicing theorists were expected to know."

For example, to give just a single illustration of this Feshbach quote, the effective range expansion was derived in one of Schwinger's 1947 lectures [26], but only published by him in 1950 [27] after Bethe published his own derivation in 1949 [28]. As Schiff [29] indicates, the importance of the effective range expansion is its demonstration that the two-nucleon scattering cross section at moderately low energies can be characterized by two lengths and two lengths alone, the scattering length and the effective range. All potentials which reproduce those two lengths are equally valid descriptions of the interaction insofar as moderately low energy nuclear scattering is concerned, irrespective of the shapes or strengths of the potentials. This inevitable conclusion of the effective range expansion is one that Breit, in his years and years of painstaking calculations which I mentioned earlier, apparently never appreciated.

I am not going to say a great deal about Schwinger's contributions (derived while he was at Harvard) to the development of modern QED that earned him the Nobel Prize. These contributions are well known, and are fully described (much better than I possibly could do here) in the publications listed in references 1 and 2. I will stress that Schwinger was the first person to develop the relativistically covariant, gauge-invariant formulation of quantum electrodynamics needed to properly incorporate renormalization into QED and thereby to correctly calculate the Lamb shift, which calculation indeed was first correctly performed by Schwinger. Schwinger also was the first person to realize that the same relativistically covariant, gauge-invariant renormalization technique that accounted for the Lamb shift would account for the departure of the gyromagnetic ratio of the electron from 2. Schweber [30] describes Schwinger's Lamb shift calculations as follows:

"The notes of Schwinger's calculations are extant. They give proof of his awesome computational powers. Starting from the Hamiltonian for the coupled field system, Schwinger proceeded to make a canonical transformation to eliminate virtual effects to lowest order. The self-energy terms were identified and a mass renormalization was then performed...What previously had been pieces of a theory became welded and unified into a consistent and coherent quantum electrodynamics to order alpha [the fine structure constant]...To obtain the expression from which the Lamb shift can be calculated, lengthy computations involving properties of solutions of the Dirac equation, traces over photon polarizations, and integrations over photon energies had to be performed. These were carried out fearlessly and seemingly effortlessly. Schwinger just plowed ahead. Often, involved steps were carried out mentally and the answer was written down. And most important, the lengthy calculations are error free!"

A few years after these triumphs of Schwinger's, Feynman announced his ability to do the same QED calculations far more easily via his diagrammatic techniques, and Schwinger's position at the crest of theoretical physics began to recede. In the decade or so from the mid-40s to the mid-50s, however, I don't believe his preeminence really was disputed, not even by Feynman's admirers. Stanley Deser, in his talk at Schwinger's UCLA Memorial Tribute [31] recalls a 50s paper in the "Journal of Jocular Physics"—a paper "purporting to be a sure fire template for writing successful papers, guaranteed to pass any referee" that he saw at the Niels Bohr Institute:

"It began with something like 'According to Schwinger', continued with 'as conveyed by the Green's function expression' and so on--only a few blanks were left to be filled in from one's favorite work by Julian. What's more, we all know that many a serious paper along these lines is indeed to be found in the literature."

## CONCLUSION

Schwinger left Harvard for UCLA in 1972. Starting in 1966, he devoted his major research efforts to "source theory", a reformulation of QED and other field theories which allowed him to avoid the necessity of introducing renormalization to eliminate the infinities that appear in the conventional field theories [32]. However the world did not follow Schwinger on this source theory path, and he ended his teaching days rather isolated from the physics community that had been idolizing him only a few years earlier. This writer speculates that Julian chose to follow this lonely path because he chose to take what he thought was an important new route rather than explore new trails set by others. As David Saxon said in his Schwinger obituary [4]: "He, accustomed to leading, chose not to follow."

What, aside from his acknowledged achievements in QED, are likely to be Schwinger's most enduring contributions to physics? Paul Martin, another of Schwinger's Ph.D. students, lists [33]: "(i) effective range theory; (ii) scattering theory; (iii) tensor forces and quadrupole moments; (iv) variational principles; (v) [the use of] Green's functions for classical and quantum fields; (vi) [the theory of] angular momentum in terms of oscillators; (vii) Coulomb states in momentum space and oscillators in external fields; (viii) commutators of currents, including the momentum current; and (ix) [the theory of] magnetic monopoles, higher spin particles, and gravitons." And, as Martin says, "Others will add to this list."

Martin indeed has listed many enduring contributions to physics, contained in Schwinger's nearly 200 published papers, several books and a large body of unpublished work. Many physicists, however (including this writer), think Schwinger's most enduring achievements may be his students. Schwinger had at least 73 students who earned their Ph.D.s with him [34]. Furthermore quite a number of these students have had distinguished productive physics careers, e.g., to name just a few such: Kenneth Case, Lowell Brown, Bryce DeWitt, Abraham Klein, Eugen Merzbacher, Roger Newton and Fritz Rohrlich [35]. In fact four of Schwinger's students are Nobel Laureates: Ben Mottelson, Shelly Glashow, Roy Glauber and Walter Kohn (in chemistry).

In short, even if explicitly mentioning Schwinger's name became unfashionable, these Ph.D. students produced by Schwinger continued (and many still continue) to instill Schwinger's techniques and his ways of approaching physics problems into the psyches of new generations of physics students. I will close these Memories, therefore, with words Saxon wrote in 1994 [4], "Through his students, he has had a more widespread and profound influence on theoretical physics over the past forty years than any other physicist."

Leventhal 2008), pp. 125-128.  Page 133 of the World Scientific publication describes Schwinger's first seminar in Oppenheimer's group at Berkeley, as well as Oppenheimer's reaction thereto (discussed below at the end of these Memories' subsection titled "SCHWINGER AS J. ROBERT OPPENHEIMER'S POSTDOC").

8.  "Remembering Julian Schwinger," APS Forum on the History of Physics Newsletter, vol. 9 No. 5, pp. 23-24 Fall 2005).  This FHP Newsletter article was the subject of a Letter to the Editor by Hans Christian von Bayer, with a response by this writer, in FHP Newsletter vol. 9 No. 6, p. 19 (Spring 2006).

9.  Some of the assertions about Schwinger made below can be found (briefly presented) in the two publications identified in reference 1 above, based (as those publications state) on telephone interviews with this writer.

10.  Indeed my own University of Pittsburgh physics department has posted this Panel, as have a number of other physics departments where I have given my Recollections of Oppenheimer and Schwinger colloquium talk.

11.  Richard Feynman, in a talk he gave at Schwinger's 60th birthday celebration.  Feynman was describing a visit by Schwinger to Los Alamos in 1943, right after the Trinity test.

12.  "He once remarked, by way of explanation, that he had been reading the family *Encyclopedia Britannica* straight through.  When he came to the letter P and physics, that did it,"  a quote (very slightly modified) from David Saxon, "Julian Schwinger Memorial Tribute," in Jack Ng's collection of talks, p. 80 (see reference 3 above).  David Saxon became a lifelong friend of Schwinger's after meeting him in 1943, when both were working at the WWII MIT Radiation Laboratory.

13.  Quoting from p. 276 of Schweber (reference 2 above).

14. "On the polarization of electrons by double scattering" (with O. Halpern, *Phys. Rev.* **48**, 109 (1935); "On the β-radioactivity of neutrons" (with L. Motz), *Phys. Rev.* **48**, 704 (1935).

15. "On the magnetic scattering of neutrons" *Phys. Rev.* **51**, 544; "On Non-Adiabatic Processes in Inhomogeneous Fields" *Phys. Rev.* **51**, 648; "On the Spin of the Neutron" *Phys. Rev.* **52**, 1250 (L).

16. "The Scattering of Neutrons by Ortho and Para Hydrogen" (with E. Teller) *Phys. Rev.* **51**. 775 (L) and *Phys. Rev.* **52**. 286.

17. Quoting from Schweber (ref. 2 above), p. 279.

18. Leonard I. Schiff, "Quantum Mechanics" (McGraw-Hill), whose first edition was published in 1949. Schiff later was elected to the National Academy and served as Chair of the Stanford physics department.

19. See also, e.g., Jeremy Bernstein, "Oppenheimer. Portrait Of An Enigma" (Ivan Dee 2004), pp. 184-185. Bernstein describes how the questioning by Oppie (then Director of the Princeton Institute for Advanced Study) during a pair of lectures by Freeman Dyson "reduced Dyson to silence". Dyson had come to the Institute to describe his very important researches showing that the formulations of quantum electrodynamics by Schwinger and by Feynman, though apparently so different, actually were identical. Only after Hans Bethe spoke to Oppie about the importance of Dyson's work did Oppie remain reasonably silent during Dyson's presentation, thereby permitting Dyson to complete his planned set of lectures.

20. Schweber (ref. 2 above), p. 288, as well as Mehra and Milton (ref. 1 above), p. 57 quote Rabi as saying Oppie initially was "terribly disappointed with Schwinger". Mehra and Milton ascribe this assertion of Rabi's to a talk by Rabi at Schwinger's 60[th] birthday Celebration (see Mehra and Milton's reference 4 on

32. For further details about source theory, see reference 1 above, especially Milton's 2007 publication.

33. Paul Martin, in Ng (reference 3 above), p. 83.

34. Schwinger's Ph. D. students are listed in Mehra and Milton (reference 1 above), pp. 639-643, and in Ng (reference 3 above), pp. 181-182.

35. In this regard there is an interesting contrast with Feynman. Feynman had only a handful of Ph.D. students, none of whom (to my knowledge) ever were particularly distinguished, although I hasten to say that I know I could be wrong about this and stand quite ready to be corrected.